\documentclass[final,5p,twocolumn]{elsarticle}

\usepackage{amssymb}
\usepackage{graphicx}

\usepackage{amsmath}
\usepackage[colorlinks=true,linktocpage=true,linkcolor=blue,citecolor=blue]{hyperref}

\biboptions{sort&compress}

\def\ba{\begin{eqnarray}}
\def\ea{\end{eqnarray}}
\def\lb{\label}

\def\bi{\bibitem}

\def\D{\Delta}

\def\e{\eta}

\def\l{\lambda}

\begin{document}

\begin{frontmatter}

\title{Finite size of hadrons and Bose-Einstein correlations in $pp$ collisions at 7 TeV}

\author[uj]{Andrzej Bialas}

\author[ujk]{Wojciech Florkowski}

\author[ifj]{Kacper Zalewski}

\address[uj]{M. Smoluchowski Institute of Physics, Jagellonian University, PL-30-348~Krakow, Poland}
\address[ujk]{Institute of Physics, Jan Kochanowski University, 25-406 Kielce, Poland}
\address[ifj]{Institute of Nuclear Physics, Polish Academy of Sciences, 31-342 Krak\'ow, Poland}

\date{today}

\begin{abstract}
Space-time correlations between produced particles, induced by the composite nature of hadrons, imply specific changes in the   properties of the correlation functions for identical particles.  The expected magnitude of these  effects is evaluated using the recently published  blast-wave model analysis of the data  for $pp$ collisions at $\sqrt{s}$ = 7 TeV.   
\end{abstract}


\end{frontmatter}

\begin{abstract}
Space-time correlations between produced particles, induced by the composite nature of hadrons, imply specific changes in the   properties of the correlation functions for identical particles.  The expected magnitude of these  effects is evaluated using the recently published  blast wave model analysis of the data  for $pp$ collisions at $\sqrt{s}$ = 7 TeV.   
\end{abstract}

{\bf 1.} It has been recently pointed out \cite{bz} that since hadrons produced in high-energy collisions are not point-like objects, they cannot be uncorrelated. Indeed, being composite,   hadrons cannot occupy too close  space-time points (because  at small distance the constituents of hadrons mix and  there are no separate hadrons to interfere).   Consequently, since
 the HBT experiment measures the quantum interference between  wave functions of hadrons, it cannot see hadrons which are too close to each other.  Therefore  the distribution function of the pair of hadrons  must vanish at small distances between them. 

This implies of course a correlation in space-time. As this correlation is the {\it necessary} consequence of the composite structure of hadrons (and thus it is a general property of the system) it is interesting to investigate to what extent it modifies the accepted ideas about the   quantum interference which are, usually, derived under the assumption that such correlations can be neglected \cite{review}.

 It was already shown  in \cite{bz} that such space-time correlations may be responsible for the observation that  the two-pion  Bose-Einstein correlation function  takes values below unity \cite{l3,lep2,cms}, at variance with the well-known theorem valid when the correlations are ignored \cite{review}.

In the present paper  the investigation of this phenomenon is continued, using the recently published \cite{bfzjop} analysis of the data on HBT radii, measured by the ALICE collaboration \cite{alice2}. This allows to estimate quantitatively the magnitude of the effect and to give predictions for its size in all three directions {\it long, side} and {\it out}, commonly used in discussion of the quantum interference \cite{review}.

In the next section the consequences of the space-time correlations for
the HBT correlation functions are explained. In Sections 3 and 4 the blast-wave model used for 
the quantitative estimate of the  effect is presented. The results are presented and summarized 
in the  last two sections.

{\bf 2.} 
In absence of correlations between produced hadrons, the two-particle source function is the simple product
\ba
w(p_1,p_2;x_1x_2)= w(p_1,x_1)w(p_2,x_2)  \lb{unc}
\ea
where $w(p,x)$ is the single-particle source function (Wigner function).  Consequently,
  the Bose-Einstein correlation function between the momenta of two identical particles 
 \ba
C(p_1,p_2)\equiv  \frac{ N(p_1,p_2)}{N(p_1)N(p_2)} \lb{cc}
\ea
is  given by \cite{review}
\ba
C(p_1,p_2) &=& 1+ \frac{\tilde{w }(P_{12}; Q)\tilde{w} (P_{12}; -Q)}{w(p_1)w(p_2)} \nonumber \\
&=& 1+\frac{|\tilde{w}(P_{12},Q)|^2}{w(p_1)w(p_2)} \geq 1 \lb{uc}.
\ea
Here
\ba
\tilde{w }(P_{12}; Q) &=& \int dx \;e^{iQx}w(P_{12};x), \nonumber \\
w(p) &=& \int dx \;w(p;x), 
\ea
where $P_{12}=(p_1+p_2)/2$ and $Q=p_1-p_2$.

The data from the L3 collaboration \cite{l3} and  from the CMS colllaboration \cite{cms} show that the correlation function $C(p_1,p_2)$ takes values below unity, contrary to Eq. (\ref{uc}). Thus the particles  must be correlated and we propose that this effect is due to  the composite nature of hadrons. 
 
To implement these space-time correlations, we  replace  formula (\ref{unc}) for the two-particle source function by
\ba 
&&W(p_1,p_2;x_1,x_2) = \nonumber \\
&& \hspace{0.5cm}  w(p_1;x_1)w(p_2;x_2)[1-D(x_1-x_2)],
\ea
where $D(x_1-x_2)$ is the cut-off function that satisfies the constraint $D(x_1-x_2=0)=1$ and tends to 0 at larger distances (above, let us say,  1 fm). Then, the HBT correlation function becomes
\ba 
C(P_{12},Q)=C_{\rm noncorr}(P_{12},Q)-C_{\rm corr}(p_1,p_2),   \lb{c}
\ea
where the uncorrelated part $C_{\rm noncorr}(P_{12},Q)$ is given by (\ref{uc}), while the correction due to space-time correlations reads
\ba
C_{\rm corr}=C_{\rm corr}^{(0)}+C_{corr} ^{(Q)} \lb{corrg}
\ea 
where
\ba
C_{\rm corr}^{(0)} = \frac{\int dx_1dx_2  w(p_1;x_1)w(p_2;x_2)D(x_1-x_2)}{ w(p_1)w(p_2)}, \lb{corr0}
\ea
 \ba
&&\hspace{-1.cm} C_{\rm corr} ^{(Q)} =  \nonumber \\
&&\hspace{-1.cm} \frac{\int dx_1dx_2 e^{i(x_1-x_2)Q} w(P_{12};x_1)w(P_{12};x_2)D(x_1-x_2)}{ w(p_1)w(p_2)}.
\nonumber \\ \lb{corrQ}
\ea 
One sees that the contribution from the correlation part   is negative. Moreover, since it obtains contributions from a small region of space-time, its dependence on $Q$ is much less steep than that of the uncorrelated part. Consequently, at $Q$ large enough $C(P_{12},Q)$ may easily fall below one.

To describe the actual measurements one has to take into account that particles produced very far from the center (e.g. those arising 
from long-lived resonances) form a "halo" and  do not contribute to the HBT correlations \cite{core}. Thus we have

\ba
\hat{C}_{obs}(P_{12},Q)=1-p^2+p^2C(P_{12},Q)
\ea
where $p^2$ is the probability that both particles originate from the "core".

In the  ALICE experiment \cite{alice2}   $\hat{C}_{\rm obs}$  was, in addition,  normalized to 1 at some $Q_0$ where the influence of quantum interference is expected to be negligible.  Thus we finally have to consider the function
\ba
C_{\rm obs}(P_{12},Q)=\frac{1-p^2+p^2C(P_{12},Q)}{1-p^2+p^2C(P_{12},Q_0)}.
\ea
Introducing the (measured) intercept parameter $\l$ by the condition
\ba
1+\l\equiv C_{\rm obs}(P_{12},Q=0)
\ea
one obtains 
\ba
&&\hspace{-1.cm} p^2= \frac{\l}{C(P_{12},Q=0)-C(P_{12},Q_0)+\l[1-C(P_{12},Q_0)]} \nonumber \\
\ea
This allows to evaluate the measured correlation function in terms of the measured intercept parameter $\l$ and the evaluated correlation function $C(P_{12},Q)$.

Note that in absence of space-time correlations we have $C(P_{12},Q=0)=2$, $C(P_{12},Q_0)=1$, and thus $p^2=\l$, as is usually assumed.

{\bf 3.} To have an idea on the magnitude of the effect we discuss,  we have used the blast-wave model described  in detail in \cite{bfzjop,bfzapp}.  In this model, at freeze-out,  hadrons are created at a fixed (longitudinal) proper time 
\ba
\tau\equiv \sqrt{t^2-z^2}=\tau_f  \lb{tf}.
\ea
 The single-particle source function (in the longitudinal c.m.s. system) becomes
\ba
w(p,x) = k_0  \cosh\eta e^{-U \cosh\eta + V \cos\phi } f(r) r dr  d\phi  d\eta
\label{FT1}
\ea
where $k_0=\sqrt{m^2+k_\perp^2}$, whereas $\eta, \phi$ and $r$ are space-time rapidity, azimuthal angle and transverse distance from the symmetry axis~\footnote{All irrelevant constants are cancelled in the definition of $w(p,x)$.}. We have also introduced the notation 
\ba
U=\beta k_0\cosh\theta;\;\;\;V=\beta k_\perp\sinh\theta,
\ea
with $T=1/\beta$ being the freeze-out temperature. Finally,  $\theta$ describes  the transverse flow by the relation
\ba
\sinh\theta= \omega r,
\ea
with $\omega$ being a parameter. The function $f(r)$ describes the transverse profile of the source.

It was shown in  \cite{bfzjop} that the model is flexible enough to describe  the  HBT radii measured by the ALICE collaboration \cite{alice2}.  The function $f(r)$ was taken in the form
\ba
f(r) \sim e^{-(r-R)^2/\delta^2}
\ea
corresponding to a "shell" of the width $\sqrt{2}\, \delta$ and  radius $R$. 

Thus the model contains 5 free parameters: $T,\; \omega,\;\tau_f,\; R$ and $\delta$, which may depend on the multiplicity of the event.  Their values, giving a good description of  the HBT radii measured in \cite{alice2},  are given in \cite{bfzjop}. 

{\bf 4.}  Since we treat particles as extended objects produced on the hyperbola  (\ref{tf}), the longitudinal distance between the two hadrons located at the space-time rapidities $\e_1,\;\e_2$ should be calculated along this curve, which yields
\ba
d_\parallel=\int_{\e_1}^{\e_2}\sqrt{dz^2-dt^2}=\tau_f (\e_2-\e_1).
\ea
In the frame where $\e_1+\e_2=0$ we also have $t_1=t_2$ and thus the  total  distance between particles is 
\ba
d^2=(x_1-x_2)^2+(y_1-y_2)^2+d_\parallel^2 \equiv d_\perp^2+d_\parallel^2.
\ea
Since this expression is invariant under  boost in the longitudinal direction, it is also valid in  the LCMS system, and thus we finally have
\ba
&&\hspace{-1.2cm}  d^2(x_1,x_2)=r_1^2+r_2^2-2r_1r_2\cos(\phi_1-\phi_2)+\tau_f^2(\e_1-\e_2)^2 .
\nonumber \\ \lb{dsq}
\ea

The correlation functions were studied using   a gaussian  cut-off function
\ba
D(x_1,x_2)=e^{-d(x_1,x_2)^2/\D^2},\lb{gs}
\ea
where $\D$ is a constant fixing the scale of the cut-off region.

 \begin{figure}[h]
\begin{center}
\includegraphics[scale=0.7]{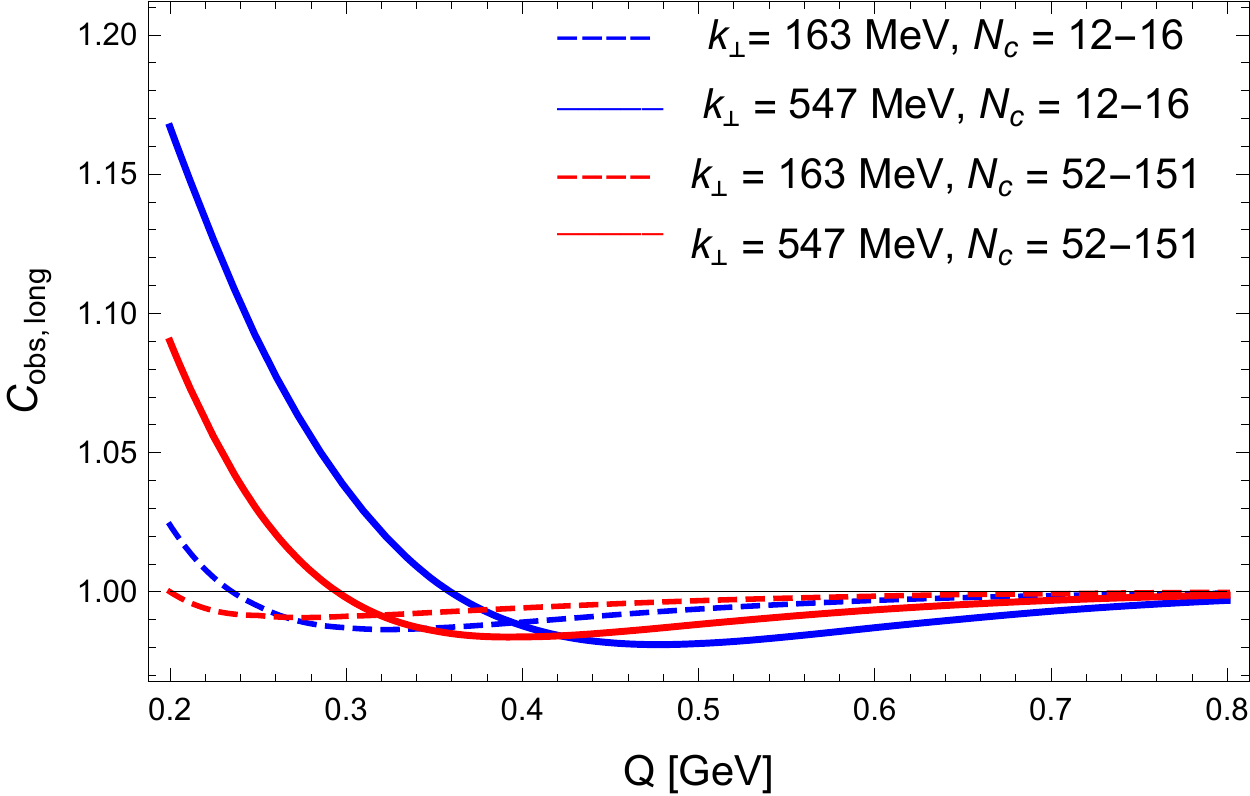}
\end{center}
\caption{(Color online) Correlation function $C_{\rm obs}$  for the $long$ direction in the interval 0.2 GeV $\leq Q \leq$ 0.8 GeV (normalized to 1 at Q= 1GeV).  The dashed lines describe the results for $k_\perp$ = 163 MeV and the two multiplicity classes: $N_c$ = 12--16 and $N_c$ = 52--151. The solid lines describe the results for $k_\perp$ = 547 MeV and the same two multiplicity classes.}%
\label{fig:long}%
\end{figure}

 \begin{figure}[h]
\begin{center}
\includegraphics[scale=0.7]{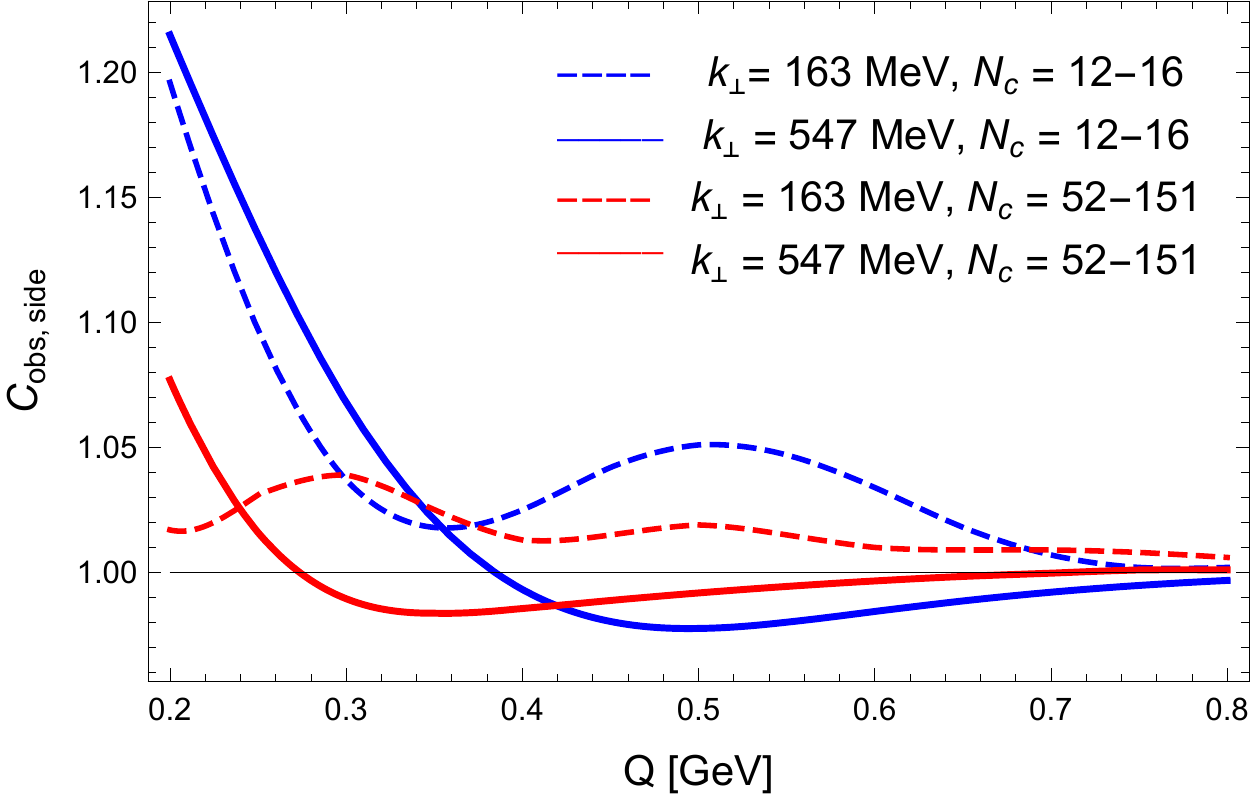}
\end{center}
\caption{ The same as Fig.~1 but for the side direction. }%
\label{fig:side}%
\end{figure}

 \begin{figure}[h]
\begin{center}
\includegraphics[scale=0.7]{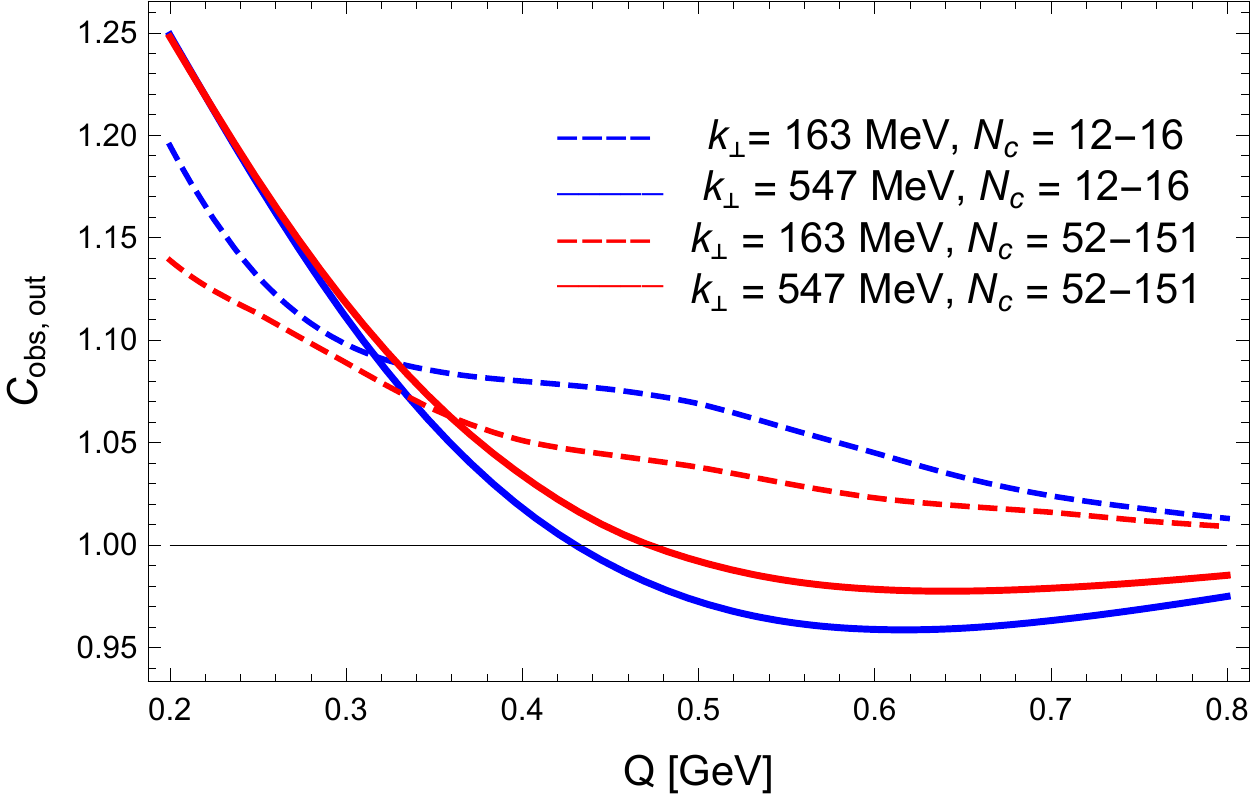}
\end{center}
\caption{ The same as Fig.~1 but for the out direction. }%
\label{fig:out}%
\end{figure}

{\bf 5.} Using  (\ref{gs}), (\ref{dsq}) and the source function of the model described in Sec. 3, with the parameters taken from \cite{bfzjop}
\footnote{ The intercept parameter $\l$ was taken as \mbox{$\l=0.59-0.26\, k_\perp$} (where $k_\perp$ is in GeV) which approximates the data of~\cite{alice2}.}, we have evaluated corrections to the HBT correlation functions  (\ref{c}) for all  intervals of the particle multiplicity and transverse momentum (as measured in \cite{alice2}), and for all three directions of the vector $\vec{Q}$. The cut-off distance  $\D\approx 2 r_V$  (where $r_V$ is  the radius of the "excluded volume"~\cite{volhist} occupied by one pion) was taken to be 1 fm,  within the range of values given by the earlier analyses \cite{vol}. Some of the results, obtained using  the gaussian $D(x_1-x_2)$, are shown in Figs.~\ref{fig:long}--\ref{fig:out}. 

One sees that for the $long$ direction the correlation function falls below 1 at all multiplicities and 
transverse momenta of the pair. The depth of the minimum in the $long$ direction varies from $\sim 0.02$ to $\sim 0.01$ (below 1) when the HBT radius $R_{\rm long}$ increases from $\sim 0.8$ to $\sim 2$ fm.

In the $side$ and $out$ directions the results are strongly dependent on the value of the transverse momentum of the pair. At $k_\perp\leq 300$ MeV for the $side$ and $k_\perp\leq 400$ MeV for the $out$ direction the correlation functions are always 
larger than 1 in the investigated region. In the $side$ direction  the correlation function shows a clear structure: a minimum followed by a maximum (particularly at  low multiplicities). At larger $k_\perp$  the minimum below 1 shows up in both cases. 

In  the  $side$ 
direction the minimum at $k_\perp\geq 300$  MeV  is similar to that found in the  $long$ direction. It is about twice deeper in  the $out$ direction (above 400 MeV). In both cases the minimum is deeper  when the  multiplicity increases. Also in this case the change is controlled by the corresponding HBT radii.

To see the sensitivity of these results to the shape of the cut-off function $D(x_1-x_2)$
we have also considered a sharp cut-off which is drastically different from the Gaussian.  We have found that the qualitative features are unchanged, except that the effects of the cut extend to larger values of $Q$. This, however, happens in the region where these effects are already  small and  rather hard to measure. Actually, in most cases the results are almost identical \footnote{In the relevant region $Q > 300$ MeV  they differ  by less than 0.01, which is consistent with the expected  accuracy of our calculations and also with the present experimental accuracy.}, provided that the cut-off parameter is   $\sim 0.75$ fm. The only exception is the $side$ direction at small multiplicity where the difference exceeds slightly 0.02 at $Q > 700$ MeV.

We thus conclude that although  the shape of the cut-off function can influence the details of our results, the general qualitative features remain unchanged.

{\bf 6.} In summary, we have estimated to what extent the space-time correlations implied by the excluded volume effect modify the HBT correlation functions.  

Our conclusions can be formulated as follows:

(i) The space-time correlations induced by the finite size of hadrons lead to a rich structure of the HBT correlation functions, depending on (i) the measurement direction, (ii) multiplicity  
and  (iii) the transverse momentum of the pair.

(ii) The difference between the $long$  and the two other  directions at small $k_\perp$ is particularly striking.

(iii) At large $k_\perp$ the minimum below 1 shows  up in every direction. It  is about twice deeper  for $out$ than for the $long$ and $side$  directions.

Some comments are in order.

(i) We have found that the modification of the HBT correlation functions  are only marginally  sensitive to the change of shape of the cut-off function $D(x_1-x_2)$. This means that the effect we discuss is, in practice,  described by a single parameter $\Delta$. 

(ii) We have been considering the space-time correlations in the source function of  two pions, which are a necessary consequence of their composite nature. Naturally, there might be also  other mechanisms contributing to these correlations (e.g. the final state interaction).  In this case the parameter $\Delta$ should be considered as an effective cut-off distance which summarizes all contributions. Since our calculations show that the measurable effects on the HBT correlation functions depend mostly  on $\Delta$ (and  not on the shape of the function $D$) it seems hopeless to try to separate the various contributions.

(iii) In our approach the cut-off function is taken independent of particle density. This approximation seems reasonable because,  as shown in \cite{bfzjop},  particle density at freeze-out changes only by $~10\%$ in the range of multiplicities  we consider. Moreover,  the dominant effect of the changing particle density  is expected to be a modification  of the  single-particle source functions  of the two pions contributing to interference rather than of their space-time correlation described by $D(x_1-x_2)$. It follows that  the observable effect of  the modification of the cut-off function due to change of particle density is expected to be very small, if any.

(iv)  It is interesting to speculate about the size of the effects we discuss in case of heavy ion collisions. Taking the source functions in the transverse direction in form of  Gaussians (which is a reasonable approximation for heavy ion collisions) one can easily see that the corrections due to  finite size of hadrons fall as $(\D/R)^2$ where $R$ is the radius of the system.  For $PbPb$ collisions this gives factor $\sim 1/30$ compared to the results shown in this paper, implying that the expected effects are negligible. Similar mechanism should be at work in the longitudinal direction. For smaller systems, as those created in $p-Pb$ collisions, the effects  are also expected to be smaller than in $pp$. Precise estimate would, however,  require  determination of the source functions  (see \cite{bfzjop}).

\bigskip

{\bf Acknowledgements}

We  thank Adam Kisiel for useful discussions and for  help in understanding the ALICE data. Thanks are also due to Viktor Begun for  an instructive discussion on the excluded volume effects. This investigation was supported in part by Polish National Science Center grant with decisions No.  DEC-2013/09/B/ST2/00497 (A.B. and K.Z.) and DEC-2012/06/A/ST2/00390 (W.F).

\end{document}